\documentclass[11pt]{article}
\usepackage{amssymb,latexsym,amsmath,amsbsy,amsthm}
\usepackage[dvips]{graphicx}
\usepackage{cite}
\usepackage{stmaryrd}  
\usepackage{arydshln}  
\usepackage{mathdots}  
\usepackage{xcolor}
\newtheorem{theo}{Theorem}
\newtheorem{defi}[theo]{Definition}

\newtheorem{prop}[theo]{Proposition}
\def\nn{\nonumber}

\def\qdots{\mathinner{\mkern1mu\raise1pt\vbox{\kern7pt\hbox{.}}\mkern2mu \raise4pt\hbox{.}\mkern2mu\raise7pt\hbox{.}\mkern1mu}}
\def\Z{{\mathbb Z}}

\def\B{\mathfrak{B}}
\def\gl{\mathfrak{gl}}

\def\g{\mathfrak{g}}
\def\h{\mathfrak{h}}
\def\so{\mathfrak{so}}
\def\sp{\mathfrak{sp}}
\def\osp{\mathfrak{osp}}
\def\pso{\mathfrak{pso}}
\def\B{\mathfrak{B}}
\def\lb{\llbracket}
\def\rb{\rrbracket}

\def\beq{\begin{equation}}
\def\eeq{\end{equation}}
\def\dg{\mathop{\rm dg}\nolimits}
\newcommand{\overbar}[1]{\mkern 3.0mu\overline{\mkern-3.0mu#1\mkern-3.0mu}\mkern 3.0mu}

\begin{document}
\begin{center}
{\Large \bf
The $\Z_2\times\Z_2$-graded Lie superalgebras $\pso(2n+1|2n)$\\[2mm] and 
$\pso(\infty|\infty)$, and parastatistics Fock spaces} \\[5mm]
{\bf N.I.~Stoilova}\footnote{E-mail: stoilova@inrne.bas.bg}\\[1mm] 
Institute for Nuclear Research and Nuclear Energy, Bulgarian Academy of Sciencies,\\ 
Boul.\ Tsarigradsko Chaussee 72, 1784 Sofia, Bulgaria\\[2mm] 
{\bf J.\ Van der Jeugt}\footnote{E-mail: Joris.VanderJeugt@UGent.be}\\[1mm]
Department of Applied Mathematics, Computer Science and Statistics, Ghent University,\\
Krijgslaan 281-S9, B-9000 Gent, Belgium.
\end{center}

\vskip 2 cm

\begin{abstract}
\noindent The parastatistics Fock spaces of order $p$ corresponding to an infinite number of parafermions and
parabosons with relative paraboson relations are constructed. The Fock spaces are lowest weight representations of 
the $\Z_2 \times \Z_2$-graded Lie superalgebra $\pso(\infty|\infty)$, with a basis consisting of row-stable Gelfand-Zetlin patterns.
\end{abstract}

\vskip 30 mm


\setcounter{equation}{0}
\section{Introduction} \label{sec:Introduction}%

All particles in the Standard Model are classified as either bosons or fermions   
obeying the symmetric Bose-Einstein statistics or the antisymmetric Fermi-Dirac statistics. 
Despite the success of the present theory, the Standard Model has  deficiencies 
such as the inability to explain the nature of dark matter and dark energy for example. 
In fact, quantum theory allows for the existence of infinitely many families of paraparticles, 
which obey mixed-symmetry statistics.  One of the first generalizations of quantum statistics, 
the so called paraboson and parafermion statistics, was introduced by Green~\cite{Green}  
already in 1953. Greenberg and Messiah~\cite{GM} considered mixed systems of parafermions $\bar f_{ j}^{\;\pm}$
($j,k,l\in \{1,2,\ldots\}$ and $\eta, \epsilon, \xi \in\{+,-\}\equiv\{+1,-1\}$): 
\begin{equation}
[[\bar f_{ j}^{\;\xi}, \bar f_{ k}^{\;\eta}], \bar f_{l}^{\;\epsilon}]=
|\epsilon -\eta| \delta_{kl} \bar f_{j}^{\;\xi} - |\epsilon -\xi| \delta_{jl}\bar f_{k}^{\;\eta}, 
\label{f-rels}
\end{equation}
and parabosons $\bar b_{ j}^{\;\pm}$ 
\begin{equation}
[\{ \bar b_{ j}^{\;\xi}, \bar b_{ k}^{\;\eta}\} , \bar b_{l}^{\;\epsilon}]= 
(\epsilon -\xi) \delta_{jl} \bar b_{k}^{\;\eta}  + (\epsilon -\eta) \delta_{kl}\bar b_{j}^{\;\xi},
\label{b-rels}
\end{equation}
and investigated the relative commutation relations between them. Following physical arguments 
they proved that there are two non-trivial relative commutation relations between parafermions and parabosons. 
The first of these are the so-called relative parafermion relations, determined by:
\begin{align}
&[[f_{ j}^{\xi}, f_{ k}^{\eta}], b_{l}^{\epsilon}]=0,\qquad [\{b_{ j}^{\xi}, b_{ k}^{\eta}\}, f_{l}^{\epsilon}]=0, \nn\\
&[[f_{ j}^{\xi}, b_{ k}^{\eta}], f_{l}^{\epsilon}]= -|\epsilon-\xi| \delta_{jl} b_k^{\eta}, \qquad
\{[f_{ j}^{\xi}, b_{ k}^{\eta}], b_{l}^{\epsilon}\}= (\epsilon-\eta) \delta_{kl} f_j^{\xi}.
\label{rel-pf}
\end{align}
The second are the so-called relative paraboson relations, defined by:
\begin{align}
&[[\bar f_{ j}^{\;\xi}, \bar f_{ k}^{\;\eta}], \bar b_{l}^{\;\epsilon}]=0,\qquad
 [\{\bar b_{ j}^{\;\xi}, \bar b_{ k}^{\;\eta}\}, \bar f_{l}^{\;\epsilon}]=0, \nn\\
&\{\{\bar f_{ j}^{\;\xi}, \bar b_{ k}^{\;\eta}\}, \bar f_{l}^{\;\epsilon}\}= 
|\epsilon-\xi| \delta_{jl} \bar b_k^{\;\eta}, \qquad
[\{\bar f_{ j}^{\;\xi}, \bar b_{ k}^{\;\eta}\}, \bar b_{l}^{\;\epsilon}]= (\epsilon-\eta) \delta_{kl} \bar f_j^{\;\xi}.
\label{rel-pb}
\end{align}
In the recent years parastatistics became again a field of increasing interest. 
On one side, possible applications of paraparticles were investigated: 
parabosons and parafermions were considered as 
candidates for the particles of dark matter/dark energy~\cite{EMM, Nelson, KY};  
quantum simulation of parabosons and parafermions~\cite{Alderete2017, Alderete2018, Alderete2021}
were proposed, thus giving a tool of potential use of paraparticles in designing quantum information systems
and as well as possible applications in optics~\cite{Walton2020}. 
On the other side, the algebraic structures behind the parabosons and parafermions were explored in order to construct 
the corresponding Fock spaces. 
The first result in this respect was given by Maekawa and Noguchi in~\cite{MN}
where the Fock space of order $p$ of $m$ parafermions is a particular irreducible module of $SO(2m+1)$ with basis vectors given by the $SO(2m+1)$ Gelfand-Zetlin (GZ) labels~\cite{GZ}. 
In~\cite{MN}, the relevant parafermion Fock spaces are identified, labelled by the order of statistics $p$, the Fock space vacuum is characterized as a lowest weight vector, and the state space vectors correspond to weight vectors of the $SO(2m+1)$ module. 
On the other hand, in this GZ-basis, the action of Chevalley elements of $SO(2m+1)$ is known, but it is not appropriate to determine the action of parafermion creation and annihilation operators.
The parastatistics algebra for a system of $m$ parafermions and $n$ parabosons with relative parafermion relations, 
determined by~\eqref{f-rels}, 
\eqref{b-rels} and~\eqref{rel-pf}, was identified in 1982 by
Palev~\cite{Palev1} and is the orthosymplectic Lie superalgebra $\mathfrak{osp}(2m+1|2n)$~\cite{Kac}.  
Ten years later, in 1992, in a review article~\cite{Palev1992} considering ordinary quantum statistics
and parastatistics Palev indicated that also the fields and the corresponding conjugate momenta in quantum field theory are generators of the infinite-dimensional orthosymplectic Lie superalgebra. He pointed out~\cite{Palev1992} that the state spaces of parastatistics can be constructed applying an induced module procedure.

The explicit Fock representations for  a system of $m$ parafermions and $n$ parabosons with relative parafermion relations were constructed in~\cite{SV2015} being 
certain infinite-dimensional lowest weight representations of $\osp(2m+1|2n)$. Moreover, the results were generalized 
to the case of an infinite number of parabosons and parafermions~\cite{SV2019}.
The second case, where~\eqref{f-rels} and~\eqref{b-rels} are combined with the relative paraboson relations~\eqref{rel-pb}, leads to an algebra which has received attention in a number of papers~\cite{YJ,YJ2,KA,KK,Tolstoy2014}, and is no longer a Lie superalgebra but a 
$\Z_2\times\Z_2$-graded Lie superalgebra~\cite{Tolstoy2014,SV2018}. Thus in\cite{SV2018} the explicit Fock representations for a system of $m$ parafermions and $n$ parabosons with relative paraboson relations
were constructed  being 
certain infinite-dimensional lowest weight representations of the $\Z_2\times\Z_2$-graded Lie superalgebra $\pso(2m+1|2n)$
(which is $\osp(1,2m|2n,0)$ in the notation of Tolstoy~\cite{Tolstoy2014}).
In the present paper we generalize the last results to the case of an infinite number of parabosons and parafermions with relative 
paraboson relations constructing their Fock spaces as a class of representations of the infinite rank $\Z_2\times\Z_2$-graded Lie superalgebra $\pso(\infty|\infty)$. In such a way we reach our ultimate goals, namely the description of parastatistics Fock
spaces with an infinite number of parafermions and parabosons with the two possible nontrivial relative relations between them. On the other hand the present paper is a further example of a physical application of the $\Z_2\times\Z_2$-graded Lie (super)algebras, as it was the case in a number of recent investigations~\cite{Aizawa1, Aizawa2, Aizawa3, Bruce1, Bruce2, Aizawa4, 
Aizawa5, Toppan1, Toppan2, Toppan3}.

The structure of the paper is as follows.  
In Section~\ref{sec:B} we give  a new matrix realization of the $\Z_2\times\Z_2$-graded Lie superalgebra $\pso(2n+1|2n)$
(as the known  one~\cite{Tolstoy2014,SV2018} cannot be extended to infinite-dimensional matrices).
In this new matrix realization, the operators corresponding to $n$ parafermions and $n$ parabosons are identified, and seen to generate a basis for $\pso(2n+1|2n)$.
The Fock space of order $p$ for such a set of parastatistics operators is identified as a lowest weight representation 
$\tilde V(p)$ of $\pso(2n+1|2n)$ in Section~\ref{sec:C}.
The difference with~\cite{SV2018} is that now the basis vectors of $\tilde V(p)$ are given in a form, appropriate 
to the case $n$  approaching infinity.  The actions of the parastatistics operators is also given. 
Finally in Section~\ref{sec:D},  the infinite rank $\Z_2\times\Z_2$-graded Lie superalgebra $\pso(\infty|\infty)$ is defined by means of a matrix form, consisting of certain infinite square matrices with only a finite number of nonzero entries.
The identification of $\pso(\infty|\infty)$ as the $\Z_2\times\Z_2$-graded Lie superalgebra generated by an infinite number of parafermions and parabosons (subject to particular mutual (paraboson) relations) is then rather straightforward.
Then we turn to the Fock spaces $\tilde V(p, \infty)$ of such combined systems of parafermions and parabosons.

\section{The $\Z_2\times\Z_2$-graded Lie superalgebra $\pso(2n+1|2n)$}
\setcounter{equation}{0} \label{sec:B}

In the present section we will introduce a different matrix realization (over the complex numbers) 
of the $\Z_2\times\Z_2$-graded Lie superalgebra $\pso(2n+1|2n)$ than the  one used in~\cite{SV2018}. 
The new matrix realization of $\pso(2n+1|2n)$ will allow us to extend the results and notation 
to the case of an infinite rank $\Z_2\times\Z_2$-graded algebra. For this purpose 
the rows and columns of matrices (and other objects) will be labelled both with negative and positive integers. 
If $n$ is a non-negative integer,  the  notation for ordered sets will be as follows:
\beq
[-n,n]=\{-n,\ldots,-2,-1,0,1,2,\ldots,n\}, \qquad [-n,n]^*=\{-n,\ldots,-2,-1,1,2,\ldots,n\}.
\eeq
Sometimes the minus sign of an index will be written as an overlined number, for example 
\[
[\bar{2},3]^*=\{\bar{2},\bar{1},1,2,3\}=\{-2,-1,1,2,3\} \qquad\hbox{and}\qquad 
[\bar{n},\bar{1}]=\{\bar{n},\ldots,\bar{2},\bar{1}\}=\{-n,\ldots,-2,-1\}.
\]
In addition we will use
\[
\Z^*=\Z\setminus\{0\},\qquad \Z_+=\{0,1,2,\ldots\},\qquad \Z_+^*=\{1,2,3,\ldots\}
\]
and similarly for $\Z_-$ and $\Z_-^*$.

Let $I$ and $J$ be the $(2\times 2)$-matrices
\beq
I:=\left(\begin{array}{cc} 0&1 \\ 1&0 \end{array}\right), \qquad
J:=\left(\begin{array}{cc} 0&1 \\ -1&0 \end{array}\right). 
\eeq

The $\Z_2\times\Z_2$-graded Lie superalgebra $\pso(2n+1|2n)$ can be defined as the set of all 
$(4n+1) \times (4n+1)$ block matrices $Y$ of the form
\beq
Y:=
\left( \begin{array}{ccc:c|ccc}
Y_{\bar{n},\bar{n}} & \cdots & Y_{\bar{n},\bar{1}} & Y_{\bar{n},0} & Y_{\bar{n},1} & \cdots & Y_{\bar{n},n}\\
\vdots & \ddots & \vdots & \vdots & \vdots & \ddots & \vdots \\
Y_{\bar{1},\bar{n}} & \cdots & Y_{\bar{1},\bar{1}} & Y_{\bar{1},0} & Y_{\bar{1},1} & \cdots & Y_{\bar{1},n}\\
\hdashline
Y_{{0},\bar{n}} & \cdots & Y_{{0},\bar{1}} & 0 & Y_{{0},1} & \cdots & Y_{{0},n}\\
\hline
Y_{{1},\bar{n}} & \cdots & Y_{{1},\bar{1}} & Y_{{1},0} & Y_{{1},1} & \cdots & Y_{{1},n}\\
\vdots & \ddots & \vdots & \vdots & \vdots & \ddots & \vdots \\
Y_{{n},\bar{n}} & \cdots & Y_{{n},\bar{1}} & Y_{{n},0} & Y_{{n},1} & \cdots & Y_{{n},n}
\end{array} \right),
\label{Y}
\eeq
where any matrix  $Y_{ij}$ with $i,j \in[\bar{n},n]^*$ is a $(2\times 2)$-matrix, $Y_{0,i}$ is a $(1\times 2)$-matrix,
 $Y_{i,0}$ a $(2\times 1)$-matrix and
\begin{align*}
& I Y_{\bar{i},\bar{j}}+Y_{\bar{j},\bar{i}}^T I=0, \quad
JY_{{i},{j}}+Y_{{j},{i}}^T J=0, \quad
I Y_{\bar{i},{j}}+Y_{{j},\bar{i}}^T J=0 \qquad(i,j\in[1,n]);\\
& Y_{0,\bar{j}}+Y_{\bar{j},0}^T I=0, \quad Y_{0,{j}}-Y_{{j},0}^T  J=0 \qquad(j\in[0,n]).
\end{align*}

The $\Z_2\times\Z_2$ grading for matrices of the form~\eqref{Y} is  determined by
\begin{equation}
\left(\begin{array}{c:c:c} 
\g_{(0,0)} & \g_{(1,1)} & \g_{(0,1)} \\ \hdashline
\g_{(1,1)} & 0 & \g_{(1,0)} \\ \hdashline
\g_{(0,1)} & \g_{(1,0)} & \g_{(0,0)} 
\end{array}\right).
\label{nosp-grading}
\end{equation}
It is straightforward to verify that the defining identities of a $\Z_2\times\Z_2$-graded Lie superalgebra~\cite{Rit1, Rit2} 
$\g=\bigoplus_{\boldsymbol{a}} \g_{\boldsymbol{a}} =
\g_{(0,0)} \oplus \g_{(0,1)} \oplus \g_{(1,0)} \oplus \g_{(1,1)}$ with $\boldsymbol{a}=(a_1,a_2)$ 
 an element of $\Z_2\times\Z_2$, namely
\begin{align}
& \lb x_{\boldsymbol{a}}, y_{\boldsymbol{b}} \rb \in \g_{\boldsymbol{a}+\boldsymbol{b}}, \label{grading}\\
& \lb x_{\boldsymbol{a}}, y_{\boldsymbol{b}} \rb = -(-1)^{\boldsymbol{a}\cdot\boldsymbol{b}} 
\lb y_{\boldsymbol{b}}, x_{\boldsymbol{a}} \rb, \label{symmetry}\\
& \lb x_{\boldsymbol{a}}, \lb y_{\boldsymbol{b}}, z_{\boldsymbol{c}}\rb \rb =
\lb \lb x_{\boldsymbol{a}}, y_{\boldsymbol{b}}\rb , z_{\boldsymbol{c}} \rb +
(-1)^{\boldsymbol{a}\cdot\boldsymbol{b}} \lb y_{\boldsymbol{b}}, \lb x_{\boldsymbol{a}}, z_{\boldsymbol{c}}\rb \rb,
\label{jacobi}
\end{align} 
where
\begin{equation}
\boldsymbol{a}+\boldsymbol{b}=(a_1+b_1,a_2+b_2)\in \Z_2\times\Z_2, \qquad
\boldsymbol{a}\cdot\boldsymbol{b} = a_1b_1+a_2b_2, \label{m}
\end{equation}
hold for homogeneous elements $x_{\boldsymbol{a}}, y_{\boldsymbol{b}}, z_{\boldsymbol{c}}$ of the form~\eqref{Y}, 
with the bracket given in terms of matrix multiplication:
\begin{equation}
\lb x_{\boldsymbol{a}}, y_{\boldsymbol{b}} \rb = x_{\boldsymbol{a}}\cdot y_{\boldsymbol{b}}
-(-1)^{\boldsymbol{a}\cdot\boldsymbol{b}}  y_{\boldsymbol{b}}\cdot x_{\boldsymbol{a}} \label{mm} .
\end{equation}

Let $e_{ij}$ be the matrix with zeros everywhere except a $1$ on position $(i,j)$, where the row and column indices run from $-2n$ to $2n$.
Consider the following elements  ($i\in[1,n]$):
\begin{align}
&\bar c_{-i}^{\;+}\equiv \bar f_{-i}^{\;+}= \sqrt{2}(e_{-2i,0}-e_{0,-2i+1}), \nn\\
&\bar c_{-i}^{\;-}\equiv \bar f_{-i}^{\;-}= \sqrt{2}(e_{0,-2i}-e_{-2i+1,0}), \;
\label{f-as-e}\\
&\bar c_{i}^{\;+} \equiv \bar b_{i}^{\;+}= \sqrt{2}(e_{0,2i}+e_{2i-1,0}), \nn\\
&\bar c_{i}^{\;-}\equiv \bar b_{i}^{\;-}= \sqrt{2}(e_{0,2i-1}-e_{2i,0}). \; 
\label{b-as-e}
\end{align}
These elements satisfy the triple relations~\eqref{f-rels}, 
\eqref{b-rels} and~\eqref{rel-pb} and  the algebra is generated by them.
In~\cite{Tolstoy2014}, the following was proved:
\begin{theo}[Tolstoy]
The $\Z_2\times\Z_2$-graded Lie superalgebra $\g$ defined by $4n$ generators $\bar f_k^\pm$ 
and $\bar b_k^\pm$ ($k=1,\ldots,n$),
where $\bar f_k^\pm \in\g_{(1,1)}$ and $\bar b_k^\pm \in\g_{(1,0)}$,
subject to the relations~\eqref{f-rels}, \eqref{b-rels} and~\eqref{rel-pb},
is isomorphic to $\pso(2n+1|2n)$.
\end{theo}

In terms of the parastatistics 
generators \eqref{f-as-e} and \eqref{b-as-e} the basis elements of the graded parts of $\g$ are as follows:
\begin{align*}
& \g_{(1,1)}~:\qquad \bar f_j^+,\quad \bar f_j^- \\
& \g_{(1,0)}~:\qquad \bar b_j^+,\quad \bar b_j^- \\
& \g_{(0,0)}~:\qquad [\bar f_j^\xi, \bar f_k^\eta], \quad \{\bar b_j^\xi, \bar b_k^\eta\} \\
& \g_{(0,1)}~:\qquad \{\bar f_j^\xi, \bar b_k^\eta\}.
\end{align*}

It is straightforward to check that $\g_{(0,0)}\oplus\g_{(1,1)} = \so(2n+1)\oplus\sp(2n)$. 
Since the even subalgebra of the Lie superalgebra $\osp(2n+1|2n)$ is also $\so(2n+1)\oplus\sp(2n)$,
we will refer to the diagonal matrices of~\eqref{Y} as the Cartan subalgebra $\h$ of $\pso(2n+1|2n)$. 
A basis of $\h$ is as follows:
\begin{align}
& 
h_{i} = \frac12 \lb  \bar c_i^{\;+}, \bar c_i^{\;-}\rb \quad (i\in [-n,n]^*). \label{Cartan-h}
\end{align}
In terms of the dual basis of $\h^*$, denoted by $\epsilon_i$ ($i\in [-n,n]^*$), 
 $\pso(2n+1|2n)$ has the same root space decomposition  as $\osp(2n+1|2n)$,
with the same positive and negative roots (but now graded with respect to $\Z_2\times\Z_2$ instead of $\Z_2$).
So, $\bar c_j^{\;+}$ are positive root vectors, and 
$\bar c_j^{\;-}$ are negative root vectors.

The Lie superalgebra $\gl(n|n)$ is a subalgebra of $\pso(2n+1|2n)$ and the $\gl(n|n)$  basis elements are given by: 
\begin{equation}
E_{jk} = \frac12 \lb  \bar c_j^{\;+}, \bar c_k^{\;-}\rb \quad (j,k\in [-n,n]^*), \label{E}
\end{equation}
where 
\begin{align}
& E_{jk} \;\hbox{ for } \; j,k\in [-n,-1]\; \hbox{ or } \; j,k\in [1,n] \; \hbox{ is even; so its degree ($\dg$) is $0$}, \label{evenE}\\
& E_{jk}\;\hbox{ for } \; j\in [-n,-1], k\in [1,n] \; \hbox{ or }\; j\in [1,n], k\in[-n,-1] \; \hbox{ is odd; so its degree is $1$},
\label{oddE}
\end{align}
satisfying the standard relations for the  basis elements of the Lie superalgebra $\gl(n|n)$   
\begin{equation}
E_{ij}E_{kl} - (-1)^{\dg(E_{ij})\dg(E_{kl})}E_{kl}E_{ij} =
\delta_{jk} E_{il} - (-1)^{\dg(E_{ij})\dg(E_{kl})} \delta_{il} E_{kj}.
\end{equation}
Note that (see~\eqref{Cartan-h} and \eqref{E} for $j=k$) $\h =\hbox{span}\{h_i, i\in[-n,n]^*\}$ 
is the Cartan subalgebra of both $\pso(2n+1|2n)$ and $\gl(n|n)$.

\section{Fock representations of $\pso(2n+1|2n)$ in the ``odd basis''}
\setcounter{equation}{0} \label{sec:C}

The parastatistics Fock space of order $p$ (for the relative paraboson relations), with $p$ a positive integer, has been constructed before~\cite{SV2018} as an infinite-dimensional lowest weight representation $\tilde V(p)$ 
of the algebra $\pso(2n+1|2n)$.  
By definition~\cite{GM} the parastatistics Fock space $\tilde V(p)$ is the Hilbert space with vacuum vector $|0\rangle$, 
defined by means of 
\begin{align}
& \langle 0|0\rangle=1, \qquad \bar c_{j}^{\;-} |0\rangle = 0, 
\qquad (\bar c_j^{\;\pm})^\dagger = \bar c_j^{\;\mp}, \qquad
\lb \bar c_j^{\;-}, \bar c_k^{\;+} \rb |0\rangle = p\delta_{jk}\, |0\rangle \quad
(j,k\in[\bar{n},n]^*)
\label{Fock}
\end{align}
and  is irreducible  under the action of the algebra $\pso(2n+1|2n)$ generated by the elements $\bar c_j^{\;\pm}$.

The vector $|0\rangle$ is the lowest weight vector of $\tilde V(p)$ with weight $[-\frac{p}{2},\ldots, -\frac{p}{2};\frac{p}{2},\ldots, \frac{p}{2}]$ in the basis $\{\epsilon_{-n},\ldots,\epsilon_{-1};\epsilon_1,\ldots,\epsilon_n\}$ (see~\eqref{Cartan-h}).
In~\cite{SV2018}, applying the induced module procedure, the $\pso(2n+1|2n)$ representation $\overline  V(p)$ with lowest weight
$[-\frac{p}{2},\ldots, -\frac{p}{2};\frac{p}{2},\ldots, \frac{p}{2}]$ was constructed. In general the $\overline  V(p)$ module 
is not irreducible and $\tilde V(p)$ is the quotient module 
\begin{equation}
\tilde V(p) = \overline V(p) / M(p),
\label{Vp}
\end{equation}
where  $M(p)$ is the maximal invariant submodule  of $\overline V(p)$.
Since the module $\overline V(p)$ has the same weight structure as the corresponding induced module 
for $\osp(2n+1|2n)$ its decomposition  with respect to $\gl(n|n)$ is the same, yielding all covariant $\gl(n|n)$ representations
labeled by a partition $\lambda$ with $\lambda_{n+1}\leq n$.
Thus the odd Gelfand-Zetlin basis (GZ) of $\gl(n|n)$ can be used and the difference, compared to~\cite{SV2018},
is that we must use a different $\gl(n|n)$~\cite{SV2016} GZ-basis, appropriate to generalize the results to $n$ approaching
infinity.  The union of all  GZ-bases of $\gl(n|n)$~\cite{SV2016} is  the basis for $\overline V(p)$. 
The notation of these basis vectors, as in~\cite[(3.9)-(3.10)]{SV2016}, 
is as follows:
\beq
 |p;m)^{2n} \equiv |m)^{2n} = \left| \begin{array}{l} [m]^{2n} \\[2mm] |m)^{2n-1} \end{array} \right)= \hspace{5cm} 
\label{mn}
\eeq
\begin{equation*}
 \left|
\begin{array}{llcll:llclll}
m_{\bar n,2n} & m_{\overbar{n-1},2n} & \cdots & m_{\bar 2,2n} & m_{\bar 1,2n} & m_{1,2n} & m_{2,2n} &\cdots & m_{n-2,2n} &m_{n-1,2n} &m_{n,2n}\\
 \uparrow & \uparrow & \cdots & \uparrow &\uparrow & &&&&&\\
m_{\bar n, 2n-1} & m_{\overbar{n-1}, 2n-1} & \cdots & m_{\bar 2, 2n-1} & m_{\bar 1, 2n-1} & m_{1,2n-1} & m_{2,2n-1} &\cdots &m_{n-2,2n-1} & m_{n-1,2n-1} &\\
&&&&&\downarrow & \downarrow & \cdots & \downarrow &\downarrow  \\
 & m_{\overbar{n-1},2n-2} & \cdots & m_{\bar 2,2n-2} & m_{\bar 1,2n-2} & m_{1,2n-2} & m_{2,2n-2} &\cdots & m_{n-2,2n-2} & m_{n-1,2n-2} &\\
 &\uparrow & \cdots & \uparrow &\uparrow &&&&\\
 & m_{\overbar{n-1},2n-3} & \cdots & m_{\bar 2,2n-3} & m_{\bar 1,2n-3} & m_{1,2n-3} & m_{2,2n-3} &\cdots & m_{n-2,2n-3}  &  &\\
 &  &\ddots &\vdots & \vdots & \vdots &\vdots &\iddots & & \\
&&& m_{\bar 2 4} &  m_{\bar 1 4} & m_{14} & m_{24}& & & & \\
&&&\uparrow & \uparrow \\
&&& m_{\bar 2 3} &  m_{\bar 1 3} & m_{13} && & & & \\
&&&&&\downarrow \\
&&&& m_{\bar 1 2} & m_{12} & & & & & \\
&&&& \uparrow\\
&&&& m_{\bar 1 1}  & & & & & &
\end{array}
\right)
\end{equation*}
where  $m_{i,2n}\in\Z_+$ are fixed and
\begin{equation}
 \begin{array}{rl}
1.& m_{j,2n}-m_{j+1,2n}\in{\mathbb Z}_+ , \;j\in[\bar{n},\bar{2}]\cup[1,n] \hbox{ and }
     m_{-1,2n}\geq \# \{i:m_{i,2n}>0,\; i\in[1,n] \};\\
2.& m_{-i,2s}-m_{-i,2s-1}\equiv\theta_{-i,2s-1}\in\{0,1\},\quad 1\leq i \leq s\leq n ;\\
3.& m_{i,2s}-m_{i,2s+1}\equiv\theta_{i,2s}\in\{0,1\},\quad 1\leq i\leq s\leq n-1 ;\\    
4.& m_{-1,2s}\geq \# \{i:m_{i,2s}>0,\; i\in[1,s] \}, \; s\in[1,n] ;\\  
5.& m_{-1,2s-1}\geq \# \{i:m_{i,2s-1}>0,\; i\in[1,s-1] \}, \; s\in[2,n] ;\\ 
6.& m_{i,2s}-m_{i,2s-1}\in{\mathbb Z}_+\hbox{ and } m_{i,2s-1}-m_{i+1,2s}\in{\mathbb Z}_+,\quad 1\leq i\leq s-1\leq n-1;\\
7.& m_{-i-1,2s+1}-m_{-i,2s}\in{\mathbb Z}_+\hbox{ and } m_{-i,2s}-m_{-i,2s+1}\in{\mathbb Z}_+,\quad 1\leq i\leq s\leq n-1.
 \end{array}
\label{cond3}
\end{equation}
Under the adjoint action in $\pso(2n+1|2n)$ of the $\gl(n|n)$ basis elements $E_{jk}$, 
\eqref{E}, the ordered set
$(\bar c_n^{\;+}, \bar c_{-n}^{\;+},\ldots,\bar c_{2}^{\;+}, \bar c_{-2}^{\;+}, \bar c_{1}^{\;+}, \bar c_{-1}^{\;+})$ is  a standard $\gl(n|n)$ tensor of rank $(1,0,\ldots,0)$ and  these $2n$  elements correspond, in this order, to 
 a unique GZ-pattern with $k$ top lines
$1 0 \cdots 0$  and $2n-k$ bottom rows of the form $0 \cdots 0$ for $k=1,2,\ldots,2n$. 
It is convenient to introduce a notation for the order in which these $2n$ elements appear:
\beq
\rho(i)= \left\{ \begin{array}{rcl}
 {2i} &  \hbox{for} &  i\in[1,n]  \\ 
 {-2i-1} &  \hbox{for} &  i\in[\bar{n},\bar{1}]
 \end{array}\right. .
\label{rho}
\eeq
Then the pattern corresponding to $\bar c^{\;+}_i$ has rows of the form $1 0 \cdots 0$ for each row index $j\in[\rho(i),2n]$ and zero rows for each row index $j\in[1,\rho(i)-1]$.

If $W([m]^{2n})$ is the $\gl(n|n)$ module with highest weight $[m]^{2n}$, 
the tensor product rule for covariant representations of $\gl(n|n)$ reads~\cite{King1990}:
\begin{equation}
W([1,0,\ldots,0]) \otimes W([m]^{2n}) = \bigoplus_{k\in[-n,n]^*} W([m]_{+(k)}^{2n}), \label{tensprod}
\end{equation}
where  $[m]_{\pm(k)}^{2n}$ is obtained from $[m]^{2n}$ by the replacement of $m_{k,2n}$ by $m_{k,2n}\pm 1$.
On the right hand side of~\eqref{tensprod} the summands for which the conditions~1. in \eqref{cond3}  are not fulfilled are omitted.

The matrix elements of $\bar c_i^{\;+}$ in $\overline V(p)$ can be written as follows~\cite{Vilenkin}:
\begin{align}
{}^{2n}(m' | \bar c_i^{\;+} | m )^{2n} & = 
\left( \begin{array}{ll} [m]^{2n}_{+(k)} \\[1mm] |m')^{2n-1} \end{array} \right| \bar c_i^{\;+}
\left| \begin{array}{ll} [m]^{2n} \\[1mm] |m)^{2n-1} \end{array} \right) \nn\\
& = \left( 
\begin{array}{c}1 0 \cdots 0 0\\[-1mm] 1 0 \cdots 0\\[-1mm] \cdots\\[-1mm] 0 \end{array} ;
\begin{array}{ll} [m]^{2n} \\[2mm] |m)^{2n-1} \end{array}  \right.
\left| 
\begin{array}{ll} [m]^{2n}_{+(k)} \\[2mm] |m')^{2n-1} \end{array} \right)
\times
([m]^{2n}_{+(k)}||\bar c^{\;+}||[m]^{2n}).
\label{mmatrix}
\end{align}
The first factor in the right hand side of~\eqref{mmatrix} is a $\gl(n|n)$ Clebsch-Gordan coefficient (CGC) 
determined in~\cite{SV2019, SV2019a}
and the second factor in~\eqref{mmatrix} is a {\em reduced matrix element} for the standard representation.
The  values of the patterns $|m')^{2n}$ are determined by the $\gl(n|n)$ tensor product rule~\eqref{tensprod} 
and the first line of $|m')^{2n}$
is of the form $[m]^{2n}_{+(k)}$. The reduced matrix elements depend only on the $\gl(n|n)$ highest weights
 $[m]^{2n}$ and $[m]^{2n}_{+k}$ (and not on the  GZ basis).
They have been determined in~\cite[formulas (3.18) and (A.4-A.7)]{SV2018}:
\begin{align}
& ([m]^{2n}_{+(k)}||\bar c^{\;+}||[m]^{2n}) = 
\tilde G_{n+k+1}(m_{-n,2n},m_{-n+1,2n},\ldots,m_{-1,2n},m_{1,2n},\ldots,m_{2n,2n}),\quad (k\in[-n,-1]) \label{3.8}\\
& ([m]^{2n}_{+(k)}||\bar c^{\; +}||[m]^{2n}) = 
\tilde G_{n+k}(m_{-n,2n},m_{-n+1,2n},\ldots,m_{-1,2n},m_{1,2n},\ldots,m_{2n,2n}),\quad (k\in[1,n]). \label{3.9}
\end{align}
For the matrix elements of $\bar c_i^{\;-}$, we use the Hermiticity requirement~\eqref{Fock}, 
\beq
{}^{2n}(m'|\bar c^{\;-}_i|m)^{2n} = {}^{2n}(m|\bar c^{\;+}_i|m')^{2n}.
\label{herm}
\eeq
In this way we obtain explicit actions of the $\pso(2n+1|2n)$ generators $\overline c_i^{\;\pm}$ on a basis of $\bar V(p)$:
\begin{align}
\bar c_i^{\;+}|m)^{2n} & = \sum_{k,m'} 
\left( 
\begin{array}{c}1 0 \cdots 0 0\\[-1mm] 1 0 \cdots 0\\[-1mm] \cdots\\[-1mm] 0 \end{array} ;
\begin{array}{ll} [m]^{2n} \\[2mm] |m)^{2n-1} \end{array}  \right.
\left| 
\begin{array}{ll} [m]^{2n}_{+(k)} \\[2mm] |m')^{2n-1} \end{array} \right)
([m]^{2n}_{+(k)}||\bar c^{\;+}||[m]^{2n}) 
\left| \begin{array}{ll} [m]^{2n}_{+(k)} \\[1mm] |m')^{2n-1} \end{array} \right), \label{cj+r}\\
\bar c_i^{\;-}|m)^{2n} & = \sum_{k,m'} 
\left( 
\begin{array}{c}1 0 \cdots 0 0\\[-1mm] 1 0 \cdots 0\\[-1mm] \cdots\\[-1mm] 0 \end{array} ;
\begin{array}{ll} [m]^{2n}_{-(k)} \\[2mm] |m')^{2n-1} \end{array}  \right.
\left| 
\begin{array}{ll} [m]^{2n} \\[2mm] |m)^{2n-1} \end{array} \right)
([m]^{2n}||\bar c^{\;+}||[m]^{2n}_{-(k)}) 
\left| \begin{array}{ll} [m]^{2n}_{-(k)} \\[1mm] |m')^{2n-1} \end{array} \right). \label{cj-r}
\end{align}
From the reduced matrix elements one determines the structure of the $\pso(2n+1|2n)$ irreducible
 $\tilde V(p)$ module.
All vectors $|m)^{2n}$~\eqref{mn} with $m_{\bar n,2n}\leq p$ satisfying conditions~\eqref{cond3}
constitute the basis of the irreducible 
$\pso(2n+1|2n)$ representation
 $\tilde{V} (p)$.

\begin{defi} 
A basis vector, $|m)^{2n}$ is said to be {\em row-stable} with {\em stability index}  $s$ 
if there exists a partition $\nu$ such that  rows $s,s+1,\ldots,2n$ are of the form
\[
[\nu_1,\nu_2,\ldots,0;0,0,\ldots].
\]
\end{defi} 

Note that:
\begin{enumerate}

\item The action~\eqref{cj+r} of $\bar c_i^{\;+}$ on $|m)^{2n}$ gives vectors $|m')^{2n}$ such that rows $1,2,\ldots,\rho(i)-1$ of $|m')^{2n}$ are the same as those of $|m)^{2n}$ 
and in rows $\rho(i),\ldots,2n$ there is a change by one unit for just one particular column index $r$:
$[m']^j = [m]^j+[0,\ldots,0,1,0,\ldots,0]$ for $j\in[\rho(i),2n]$.

\item The action ($k<n$) 
\beq
\bar c^{\;+}_{i_k} \cdots \bar c^{\;+}_{i_2} \bar c^{\;+}_{i_1} |0\rangle 
\label{consec-action}
\eeq
produces row-stable patterns if $n$ is sufficiently large (each $i_r\in[\bar{n},n]^*$). 

\item If $|m)^{2n}$ is row-stable with respect to row $s<2n-1$ 
the vectors $|m')^{2n}$ appearing in $\bar c^{\;+}_i |m)^{2n}$ are row-stable with respect to row $\max\{s+2,\rho(i)+1\}$.

\item Row-stable patterns remain row-stable under the action of $\bar c^{\;-}_i$'s for the same stability index.

\item If the top row of $|m)^{2n}$ has the zero partition as second part, i.e.\ it is of the form
\[
[m]^{2n}=[\nu_1,\nu_2,\ldots;0,\ldots,0]
\]
with $\nu$ a partition one can
define a map $\phi_{2n,+2}$ from the set of GZ-basis vectors $|m)^{2n}$ with zero second part 
to the set of GZ-basis vectors $|m)^{2n+2}$ with stability index $2n$ by:
\begin{align}
& |m)^{2n+2}=\phi_{2n,+2}\left( |m)^{2n} \right),\hbox{ where }\label{phi}\\
& [m]^{2n+1}=[\nu_1,\nu_2,\ldots,0,0;0,\ldots,0],\quad [m]^{2n+2}=[\nu_1,\nu_2,\ldots,0,0;0,\ldots,0,0] \nn
\end{align}
and extend it by linearity, on a linear combination of vectors $|m)^{2n}$ with zero second part.

\item Let $|m)^{2n}$ be row-stable with respect to row~$2n$, and $|m)^{2n+2}=\phi_{2n,+2}\left( |m)^{2n} \right)$.
Then for all $i$ with $\rho(i)\leq 2n$ (or equivalently, $i\in[-n,n]^*$):
\[
\bar c^{\;+}_i |m)^{2n+2} = \phi_{2n,+2}\left(\bar c^{\;+}_i|m)^{2n}\right).
\]
\end{enumerate}

\section{The Fock representations $\tilde V(p, \infty)$ of $\pso(\infty|\infty)$}
\label{sec:D}

In this Section we  extend the $\Z_2 \times \Z_2$-graded Lie superalgebra $\pso(2n+1|2n)$ and its Fock representations 
$\tilde V(p)$ to the infinite rank case $\pso(\infty|\infty)$.

Consider the set of all squared infinite matrices of the form
\beq
Y:=
\left( \begin{array}{ccc:c|ccc}
\ddots & \vdots & \vdots & \vdots & \vdots & \vdots & \iddots \\
\cdots & Y_{\bar{2},\bar{2}} & Y_{\bar{2},\bar{1}} & Y_{\bar{2},0} & Y_{\bar{2},1} & Y_{\bar{2},2} & \cdots\\
\cdots & Y_{\bar{1},\bar{2}} & Y_{\bar{1},\bar{1}} & Y_{\bar{1},0} & Y_{\bar{1},1} & Y_{\bar{1},2} & \cdots\\
\hdashline
\cdots & Y_{{0},\bar{2}}  & Y_{{0},\bar{1}} & 0 & Y_{{0},1} & Y_{{0},2}& \cdots \\
\hline
\cdots & Y_{{1},\bar{2}} & Y_{{1},\bar{1}} & Y_{{1},0} & Y_{{1},1} & Y_{{1},2}& \cdots \\
\cdots & Y_{{2},\bar{2}} & Y_{{2},\bar{1}} & Y_{{2},0} & Y_{{2},1} & Y_{{2},2}& \cdots \\
\iddots & \dots & \vdots & \vdots & \vdots & \vdots & \ddots 
\end{array} \right).
\label{Yi}
\eeq
with indices  in the set $\Z$, $Y_{ij}$ with $i,j \in\Z^*$  a $(2\times 2)$-matrix, $Y_{0,i}$  a $(1\times 2)$-matrix 
and $Y_{i,0}$ a $(2\times 1)$-matrix.
The infinite-dimensional $\Z_2 \times \Z_2$-graded Lie superalgebra $\pso(\infty|\infty)$ can be defined as the set of all squared infinite matrices of the form~\eqref{Yi} such that each matrix has only a finite number of nonzero entries, and such that the (non-zero) blocks satisfy
\begin{align*}
& I Y_{\bar{i},\bar{j}}+Y_{\bar{j},\bar{i}}^T I=0, \quad
JY_{{i},{j}}+Y_{{j},{i}}^T J=0, \quad
I Y_{\bar{i},{j}}+Y_{{j},\bar{i}}^T J=0 \qquad(i,j\in\Z_+^*);\\
& Y_{0,\bar{j}}+Y_{\bar{j},0}^T I=0, \quad Y_{0,{j}}-Y_{{j},0}^T  J=0 \qquad(j\in\Z_+).
\end{align*}
The $\Z_2 \times \Z_2$ grading for the matrices~\eqref{Yi} is determined by~\eqref{nosp-grading}
and for homogeneous elements $x_{\boldsymbol{a}}, y_{\boldsymbol{b}}$ of the form~\eqref{Yi}, 
the bracket is defined as follows:
\begin{equation}
\lb x_{\boldsymbol{a}}, y_{\boldsymbol{b}} \rb = x_{\boldsymbol{a}}\cdot y_{\boldsymbol{b}}
-(-1)^{\boldsymbol{a}\cdot\boldsymbol{b}}  y_{\boldsymbol{b}}\cdot x_{\boldsymbol{a}} .
\end{equation}
and extended by linearity (see\eqref{m} and \eqref{mm}).

The matrices $e_{ij}$ consist of zeros everywhere except a $1$ on position $(i,j)$, where the row and column indices belong to $\Z$. 
A basis of a Cartan subalgebra $\h$ of $\pso(\infty|\infty)$ consists of the elements $h_i=e_{2i-1,2i-1}-e_{2i,2i}$ ($i\in\Z_+^*$) and
$h_{i}=e_{2i,2i}-e_{2i+1,2i+1}$ ($i\in\Z_-^*$). 
The corresponding dual basis of $\h^*$ is denoted by $\epsilon_i$ ($i\in\Z^*$).
As in the finite rank case, we can identify the following  root vectors  in $\g_{(1,1)}$ ($i\in\Z_+^*$):
\begin{align}
&\bar c_{-i}^{\;+}\equiv \bar f_{-i}^{\;+}= \sqrt{2}(e_{-2i,0}-e_{0,-2i+1}), \nn\\
&\bar c_{-i}^{\;-}\equiv \bar f_{-i}^{\;-}= \sqrt{2}(e_{0,-2i}-e_{-2i+1,0}), \;
\label{f-as-ei}
\end{align}
and  in $\g_{(1,0)}$ ($i\in\Z_+^*$):
\begin{align}
&\bar c_{i}^{\;+} \equiv \bar b_{i}^{\;+}= \sqrt{2}(e_{0,2i}+e_{2i-1,0}), \nn\\
&\bar c_{i}^{\;-}\equiv \bar b_{i}^{\;-}= \sqrt{2}(e_{0,2i-1}-e_{2i,0}). \; 
\label{b-as-ei}
\end{align}
The operators $\bar c_i^{\;+}$ can be chosen as positive root vectors, and the $\bar c_i^{\;-}$ as negative root vectors.

The operators $\bar c_i^{\;\pm}$ introduced here satisfy the triple relations of parastatistics. However
 now we are dealing with an infinite number of parafermions and an infinite number of parabosons, satisfying the mutual relative paraboson relations. So, the triple relations~\eqref{f-rels}, \eqref{b-rels} and \eqref{rel-pb} are satisfied with $j,k,l\in\Z^*$. In addition: as a $\Z_2 \times \Z_2$-graded
 Lie superalgebra defined by generators and relations, 
$\pso(\infty|\infty)$ is generated by the elements $\bar c_i^{\;\pm}$ ($i\in\Z^*$) subject to the relations~\eqref{f-rels}, \eqref{b-rels} and \eqref{rel-pb}.

The parastatistics Fock space of order $p$, with $p$ a positive integer, can be defined as before, and will correspond to a lowest weight representation $\tilde V(p, \infty)$ of the algebra $\pso(\infty|\infty)$.  
$\tilde V(p, \infty)$ is the Hilbert space generated by a vacuum vector $|0\rangle$ and the parastatistics creation and annihilation operators, i.e.\ subject to $\langle 0|0\rangle=1$, $\bar c_{j}^{\;-} |0\rangle = 0$, 
$(\bar c_j^{\;\pm})^\dagger = \bar c_j^{\; \mp}$,
\begin{equation}
\lb \bar c_j^{\;-}, \bar c_k^{\;+} \rb |0\rangle = p\delta_{jk}\, |0\rangle \quad (j,k\in\Z^*)
\label{Focki}
\end{equation}
and which is irreducible under the action of the algebra $\pso(\infty|\infty)$.
Clearly $|0\rangle$ is a lowest weight vector of $\tilde V(p, \infty)$ with weight $(\ldots,-\frac{p}{2}, -\frac{p}{2} | \frac{p}{2}, \frac{p}{2}, \ldots)$ in the basis $\{\ldots,\epsilon_{-2}, \epsilon_{-1}; \epsilon_{1}, \epsilon_{2}, \ldots\}$.

The basis vectors of $\tilde V(p, \infty)$  consist of  infinite row-stable GZ-patterns. These are 
 GZ-patterns with an infinite number of rows, of the type introduced in~\eqref{mn}, but such that from a certain row index $s$ all rows $s,s+1,s+2,\ldots$ are of the same form. 
The basis of $\tilde V(p, \infty)$ is  as follows:
\begin{prop}
A basis of $\tilde V(p, \infty)$ is given by all infinite row-stable GZ-patterns $|m)^{\infty}$ of the form~\eqref{mn} with $n\rightarrow\infty$
where for each $|m)^\infty$ there should exist a row index $s$ (depending on $|m)^\infty$) such that row $s$ is of the form
\[
[m]^s=[\nu_1,\nu_2,\ldots,0;0,0,\ldots]
\]
with $\nu$ a partition, all rows above $s$ are of the same form (up to extra zeros), and $\nu_1\leq p$.
Furthermore all $m_{ij}\in\Z_+$ and the usual GZ-conditions must be satisfied (for all $r\in\Z_+^*$):
\begin{equation*}
 \begin{array}{rl}
1.& m_{-i,2r}-m_{-i,2r-1}\equiv\theta_{-i,2r-1}\in\{0,1\},\quad 1\leq i \leq r;\\
2.& m_{i,2r}-m_{i,2r+1}\equiv\theta_{i,2r}\in\{0,1\},\quad 1\leq i\leq r ;\\    
3.& m_{-1,2r}\geq \# \{i:m_{i,2r}>0,\; i\in[1,r] \} ;\\  
4.& m_{-1,2r+1}\geq \# \{i:m_{i,2r+1}>0,\; i\in[1,r] \} ;\\ 
5.& m_{i,2r+2}-m_{i,2r+1}\in{\mathbb Z}_+\hbox{ and } m_{i,2r+1}-m_{i+1,2r+2}\in{\mathbb Z}_+,\quad 1\leq i\leq r;\\
6.& m_{-i-1,2r+1}-m_{-i,2r}\in{\mathbb Z}_+\hbox{ and } m_{-i,2r}-m_{-i,2r+1}\in{\mathbb Z}_+,\quad 1\leq i\leq r.
 \end{array}
\end{equation*}
\end{prop}

The process of adding an infinite number of identical rows (up to additional zeros) at the top of a finite GZ-pattern 
is as follows:
let $|m)^{2n}$ be a finite GZ-pattern of type~\eqref{mn} with $2n$ rows, such that row $2n$ is of the form
$[\nu_1,\nu_2,\ldots;0,0,\ldots,0]$.
Then $\phi_{2n,\infty}\left( |m)^{2n}\right)$ is the infinite GZ-pattern consisting of the rows of $|m)^{2n}$ to which an infinite number of rows $[\nu_1,\nu_2,\ldots;0,0,\ldots,0]$ are added at the top (all identical, up to additional zeros).
Conversely, if an infinite GZ-pattern $|m)^{\infty}$ is given, which is stable with respect to row $2s$, then one can restrict the infinite pattern to a finite GZ-pattern, and 
\[
|m)^{2s} = \phi_{2s,\infty}^{-1}\left( |m)^{\infty}\right).
\]
Both maps are extended by linearity and 
 the action of $\bar c^{\; \pm}_i$ on vectors $|m)^\infty$ is defined as follows:
\begin{defi}  
Given a vector $|m)^\infty$ of $V(p)$ with stability index $2s$, and a generator $\bar c^{\;\pm}_i$.
Let $2n$ be such that $2n>\max\{2s,\rho(i)\}$.
Then 
\beq
\bar c^{\;\pm}_i |m)^\infty = \phi_{2n,\infty} \left( \bar c^{\;\pm}_i |m)^{2n} \right), \hbox{ where }
|m)^{2n} = \phi_{2n,\infty}^{-1}\left( |m)^\infty \right).
\label{c-inf}
\eeq
\end{defi}
\noindent Then
\begin{theo}
The vector space $\tilde V(p, \infty)$, with basis vectors all infinite row-stable GZ-patterns for which $\nu_1\leq p$, on which the action of the $\pso(\infty|\infty)$ generators $\bar c_i^{\;\pm}$ ($i\in\Z^*$) is defined by~\eqref{c-inf}, is an irreducible unitary Fock representation of $\pso(\infty|\infty)$.
\end{theo}

The proof follows the same steps as for the class of representations of the Lie superalgebra $\B (\infty,\infty)$ and its parastatistics Fock spaces given in~\cite{SV2019}.

\vskip 0.5 cm

This paper closes a long-standing program, namely the construction of parastatistics Fock spaces in explicit form. 
The non-trivial relative commutation relations between parafermions and parabosons lead to different algebraic structures – the Lie superalgebras $\osp(2n+1|2n)$, $\osp(\infty|\infty)$ and the $\Z_2\times\Z_2$-graded Lie superalgebras $\pso(2n+1|2n)$, $\pso(\infty|\infty)$. 
The basis vectors of the parastatistics Fock spaces as irreducible modules of the corresponding algebraic structures are particular GZ-patterns.  
It is clear that the vacuum $|0\rangle$ of the Fock spaces is just the GZ-vector with all labels zeros.
Expressing the other basis states of the Fock spaces (i.e.\ the other GZ-vectors) purely as a polynomial in creation operators acting on the vacuum remains a difficulty.
Even in the easier case of a finite number of parabosons only, this is already a challenging problem~\cite{BDV2021}.

The fact that the subalgebra structures of $\osp(2n+1|2n)$, $\osp(\infty|\infty)$ and $\pso(2n+1|2n)$, $\pso(\infty|\infty)$ are the same allowed us to use the same $\gl(n|n)$ GZ-patterns. 
Hence it is difficult to see the delicate distinction between the $\osp$-case and the $\pso$-case.
This genuine distinction is displayed in~\eqref{3.8}-\eqref{3.9} and the sign difference determined in~\cite[eq.~(3.18)]{SV2018}.
From here, it is also clear that the actions of paraoperators with relative paraboson relations $\bar{f}_i^{\;\pm}, \;\bar{b}_i^{\;\pm} $ are related to the actions of those with relative parafermion relations ${f}_i^{\;\pm}, \;{b}_i^{\;\pm} $ by means of a certain phase factor. 
This indicates that these two sets of paraoperators could be related to each other by a Klein transformation 
(see~\cite{Klein, Luders, Vasiliev1984, MMG} 
for the first ideas in this respect), 
as it is for example the case for $\gl(m|n)$ and a $\Z_2 \times \Z_2$-graded version $\gl(m_1,m_2|n_1,n_2)$~\cite{ISV2020}.

\section*{Acknowledgments}

N.I. Stoilova was supported by the Bulgarian National Science Fund, grant KP-06-N28/6, and J. Van der Jeugt was 
supported by the EOS Research Project 30889451. We thank the anonymous referees for their useful suggestions, which improved the quality of this paper.

\vskip 0.5cm
The authors would like to dedicate this paper to the memory of Prof.~Tchavdar Palev, who passed away on November 19, 2021.  Prof.~Palev was a dedicated mentor, a stimulating scientist, a splendid person and a great inspiration to all of us.

\end{document}